# Natural selection in the colloid world: Active chiral spirals

*Jie Zhang[1] and Steve Granick[2]\**

*[1]Department of Materials Science and Engineering, University of Illinois, Urbana, Illinois 61801, USA*

*[2]IBS Centre for Soft and Living Matter, UNIST, Ulsan 689-798, South Korea*

*\*e-mail: sgranick@ibs.re.kr*

## Abstract

We present a model system in which to study natural selection in the colloid world. In the assembly of active Janus particles into rotating pinwheels when mixed with trace amounts of homogeneous colloids in the presence of an AC electric field, broken symmetry in the rotation direction produces spiral, chiral shapes. Locked into a central rotation point by the centre particle, the spiral arms are found to trail rotation of the overall cluster. To achieve a steady state, the spiral arms undergo an evolutionary process to coordinate their motion. Because all the particles as segments of the pinwheel arms are self-propelled, asymmetric arm lengths are tolerated. Reconfiguration of these structures can happen in various ways and various mechanisms of this directed structural change are analyzed in detail. We introduce the concept of VIP (very important particles) to express that sustainability of active structures is most sensitive to only a few particles at strategic locations in the moving self-assembled structures.

"Active matter," by which we mean matter not at thermal equilibrium but instead with mechanisms such that it consumes energy fed into its environment, is of mounting topical interest.[1-4] Nanoparticle and colloidal systems are especially attractive experimental systems as each moving element can with proper design be imaged while simultaneously viewing the collective behavior.[5] Without the averaging that inevitably goes into scattering experiments,[6] and without the coarse-graining that goes into rheology and other ensemble-averaged measurements of system properties,[7] these systems can reveal not only the fine structure but also its larger-scale assembly, provided that the moving elements are large enough to visualize.

Here we are interested in the cases where the moving elements interact with attraction sufficiently strong to produce assembly, but sufficiently weak that the collective assembly has the opportunity to reconfigure in response to the environment. Aspects of traditional self-assembly are involved[8] but are even more complex because the need to respond to the environment can mean that dynamically-assembling systems are not governed primarily by particle-particle interactions as would be the case in traditional equilibrium self-assembly.[9-11] In this study, we are especially interested in how the system evolves towards a final steady-state. As would be the case for biological evolution, it is natural to anticipate some kind of natural selection of those structures that are most compatible with the environment.

When assemblies of particles move coherently in liquid, their interactions with the environment can generate structures that are chiral, i.e. structures that are the same except for different handedness. Examples in the natural world span the range from the mundane, as when water drains in a sink by forming a chiral vortex, to the celestial, as when galaxies rotate clockwise or anti-clockwise. Other examples abound even more in the biological world; for example, lower level organisms such as bacteria often swim in a spiral path as their bodies are never perfectly symmetric.[12] These examples raise interesting puzzles that go beyond simply observing the chirality of the trajectories, puzzles about the relationship between shape and motion, especially when the mounting interest in active material design must ultimately be grounded in coordinating the motion of body parts and the whole body.

Regarding the experimental system, this study goes beyond this laboratory's traditional interest in single-component Janus particles[13-15] and considers instead the new dynamic structures which result when one mixes Janus particles with isotropic particles, both of them in an electric field. Building upon a preliminary study presented elsewhere[16] in which the Janus particles are repulsive to each other, here we consider particles that mutually attract to form active filaments. This is a convenient system in which to explore how structure evolves. Geometrical considerations in this system generate rotating clusters assembled from an isotropic sphere at the core, surrounded by three or four arms composed of strings of self-propelled Janus particles linked head-to-tail. "Evolution" springs from the assembly, disassembly and conformation changes of structures through various "reactions" analyzed below. Long-lived structures are "naturally selected" because their collective motion is better compatible with the motion of the constituent individual particles, which themselves are active. We find that the particles bonded into such long-lived clusters continue to enjoy certain orientational degrees of freedom but not usually so much that this changes the structure or conformation of the moving collective structure.

### *Experimental section*

For easiest visualization in an optical microscope, micron-sized particles were selected for study, with an experimental configuration described elsewhere.[16, 17] Mixtures of silica spheres (usually 3 or 4 µm diameter) and half-metal coated silica Janus colloidal spheres (3 µm diameter silica coated on one hemispheres with 20 nm of titanium and then 20 nm of SiO$_2$) are selected with Janus spheres in excess in order to promote the assembly of numerous Janus spheres around a central core. The mixture is suspended in 0.1 mM NaCl aqueous solution and the particle suspension is sandwiched between two conductive transparent ITO coated coverslips and subjected to an AC electric field in the vertical direction. Density mismatch causes the silica spheres to sediment and ensures that the experimental system is 2D. When electric field is applied, the Janus particles stand up with their Janus interface parallel to the applied field. Observed in a bright field microscope, the metal appears black. Therefore, these Janus particles appear half-black-half-white in the figures described below. Typically, the electric field condition was 300 kHz, 40V/mm.

The uncoated silica particles are not propelled by the electric field, but electric field induces a dipole moment in both the uncoated silica and Janus particles. For the uncoated silica particles, the dipole is induced in the centre, whereas for Janus particles two dipoles are induced, one in each hemisphere of the particle,[18, 19] each one shifted from the centre of the sphere.[17] Calculation confirms that the interaction between metal dipole and silica dipole is attractive[16] and that adjoining Janus particles favors head-to-tail attachment between the metal side and the silica side. Similarly, the metal coated hemispheres of Janus particles are attracted to bare silica particles.

### *Chiral spirals*

The experimental system produces rotating spiral structures in which the core is a silica particle and arms of Janus particles emanate from the core, usually with 3 (for 3 µm cores) or 4 (for 4 µm cores) arms because this is how many arms can nucleate around the core. The arms are found to arrange spaced by equal distance. The structure is not like a rigid pinwheel however: Janus particles within each arm can be perturbed by others and by random fluctuations, so each arm is flexible. At the selected AC frequency it is known that Janus particles are propelled in the direction of the metal coated hemisphere owing to asymmetric induced-charge electroosmotic flow around the two hemispheres each of which has distinct polarizability, as described previously.[20] As the present system assembled with an excess of Janus particles, some free chains of Janus particles were also observed, unattached to a central core.

As shown in Fig. 1a, the synergistic reorientation of these active building blocks into a spinning structure leads to self-sustaining rotation that naturally breaks symmetry into spirals of opposite handedness and rotation direction. From a static image, the rotation direction can be predicted from the structure. Each arm is flexible because head-to-tail arrangement of Janus particles is satisfied over a large degenerate spectrum of possible mutual orientations. Thus, the same connectivity between particles allows numerous conceivable pathways of propelled motion, but only certain arrangements are self-sustaining, i.e., only certain arrangements are compatible with stability of the overall cluster.

Before the AC electric field is applied, there are no dipoles and the mixture of particles is homogeneous. After the AC electric field is switched on, we find that sometimes a disordered cluster assembles in the early stages, and then its structure adapts to become spiral. Alternatively, sometimes we find templated growth in which arms grow regularly from a smaller spiral. The pinwheel structures rotate because of the geometrical arrangement of the Janus particles and although the rotation breaks symmetry, there is no preference for clockwise or counter-clockwise motion. Since each Janus particle in each arm carries its own engine and drives itself, the arms of the pinwheel do not need to be of the same length to take on steady rotation. The particles adapt to the collective structure and to the hydrodynamic forces that the spiral encounters, during rotation, by adapting their orientations depending on where they reside in the cluster.

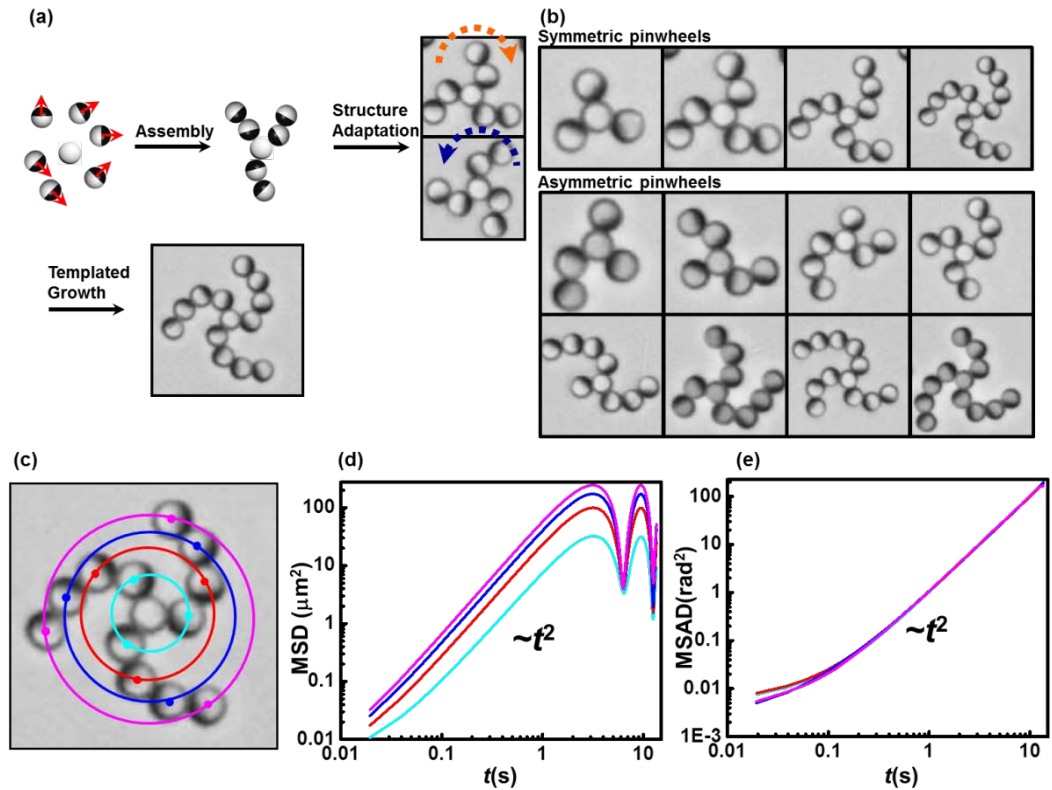

Figure 1. (a) Schematic showing the growth of chiral clusters from structure adaptation and templated growth. (b) A collection of symmetric and asymmetric chiral pinwheels (only those with clockwise motion are shown). (c) Microscopic image of a three-arm symmetric pinwheel rotating clockwise. Cyan, red, blue and magenta circles labels the position of active Janus particles in each layer and their respective trajectories. (d) Mean square displacement of active Janus particles in each layer. Colour coding is the same as in (c). (e) Mean square angular displacement of active Janus particles in each layer. Colour coding is the same as in (c) and (d).

A perfectly symmetric pinwheel structure would have its centre particle fixed in place, but in practice we find they usually jiggle around their starting position, presumably because the arm configurations fluctuate slightly. Similarly, in a perfectly symmetric structure the angular velocity of all participating particles would be the same, and the velocity would depend on distance to the centre. Fig. 1c shows the microscopic image of a pinwheel rotating clockwise with three four-particle-length arms. Cyan, red, blue, and magenta circles with dots label the trajectories and positions of the particles in each layer. All the three particles on the circles trace out identical trajectories and are rotationally the same. Fig. 1d and 1e show the mean square displacement (MSD) and mean square angular displacement (MSAD) of particles on each layer with the same colour coding as in figure 1c. Both the MSD and MSAD scale with time squared, as a result of the constant speed and angular velocity in each layer. To maintain intact structure during rotation, the angular velocity of each layer is the same in figure 1e, but the velocity in proportional to the distance of the layer to the centre, as shown in Fig. 1d.

It is striking that arms bend so that their curvature trails the motion of the rotating assembly. In general, arms trail a rotating object when the driving force is applied from the centre and transmitted to the arms, similar to the case if one places a T-shirt on a table and spins the shirt. Despite the fact that in this experimental system each particle is propelled individually, the symmetrical arrangement of arms around a central particle enforces rotation. The conformation and rotation of these pinwheels is equivalent to the case if all the arms were bonded to a centre particle that itself produced the overall rotation. It is pleasing to notice that spiral galaxies similarly display trailing arms with few exceptions.[21]

We can speculatively hypothesize however the different scenario in which the end of each arm would be driven by a tangential external force. This could be produced by a rotating magnetic field if the particles had magnetic susceptibility, as summarized in Fig. 2.

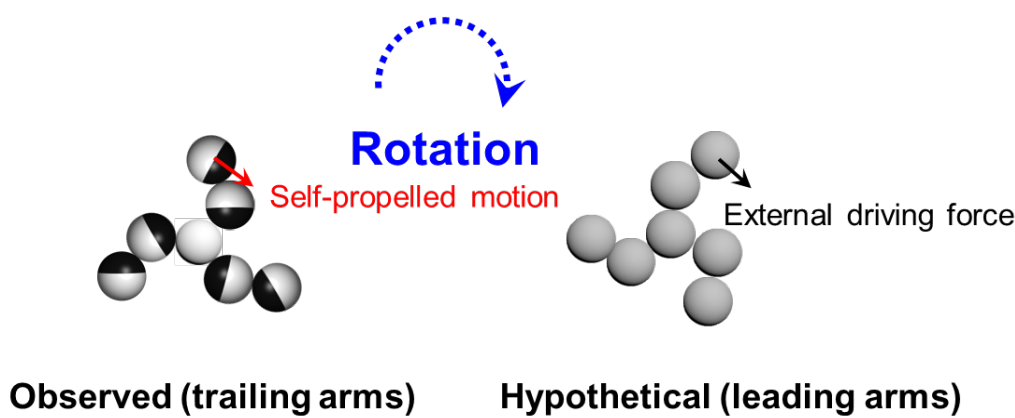

Figure 2. The trailing (experimentally observed) and leading (hypothesized) arms of a rotating pinwheel.

### *Onset of coherent motion and orientation locking*

It is a puzzle why arms lock into distinct mutual curved orientations although they are far away physically. This would be unsurprising if their mutual attachments were permanent and dictated by growth from the centre particle, but in fact the bonding between neighboring particles is fragile and sensitive to orientation of each individual particle.

In Fig. 3a, we show the assembly at 5 successive instants of time up to 3 s, of a simple three-arm pinwheel, each arm containing only one particle, starting from particles that initially are unconnected. To aid visualization, 3 of these particles are labeled with different colours. Fig. 3b shows their time-dependent orientations in the x-y plane. One observes that two particles (blue and magenta) attach to a central particle and start to rotate clockwise: the orientations of these two particles shown on figure 3b increase steadily with roughly constant speed until the third (red) particle joins. Orientation of the incoming red particle is nearly constant until it attaches but starting from about 1.4 s, its orientation abruptly changes to a new definite level, under the influence of other free particles nearby, even before attachment to the cluster.

This move happened by chance but was essential to make it possible for the red particle to approach the cluster and become attached to it. Starting from about 2 s, all the three arm particles have joined the cluster. They now move coherently with fixed orientation difference; they rotate together. Also intriguing is to note that participation of the red particle caused the blue and magenta particles to rotate backwards relative to the orientation trajectories at early times, which one observes from the microscopic images and in the unusual turns of the three curves in Fig. 3b at around 2 s elapsed time. It seems that the last member to join played the most important role in determining the configuration of the final cluster.

Figure 3c illustrates a pinwheel with three longer arms. The cluster first translates in space because the pinwheel was asymmetric in arm length and arm orientation; motion of the entire cluster is incompatible with that of its individual constituents. This causes the cluster to deform and gradually modulates its shape, tending towards a sustainable shape. In the final state the centre no longer translates apart from fluctuations. It is interesting that before arriving at the final state, the arms seem to be flexible and subject to perturbation, in contrast to the final state in which the arms become reasonably static in shape.

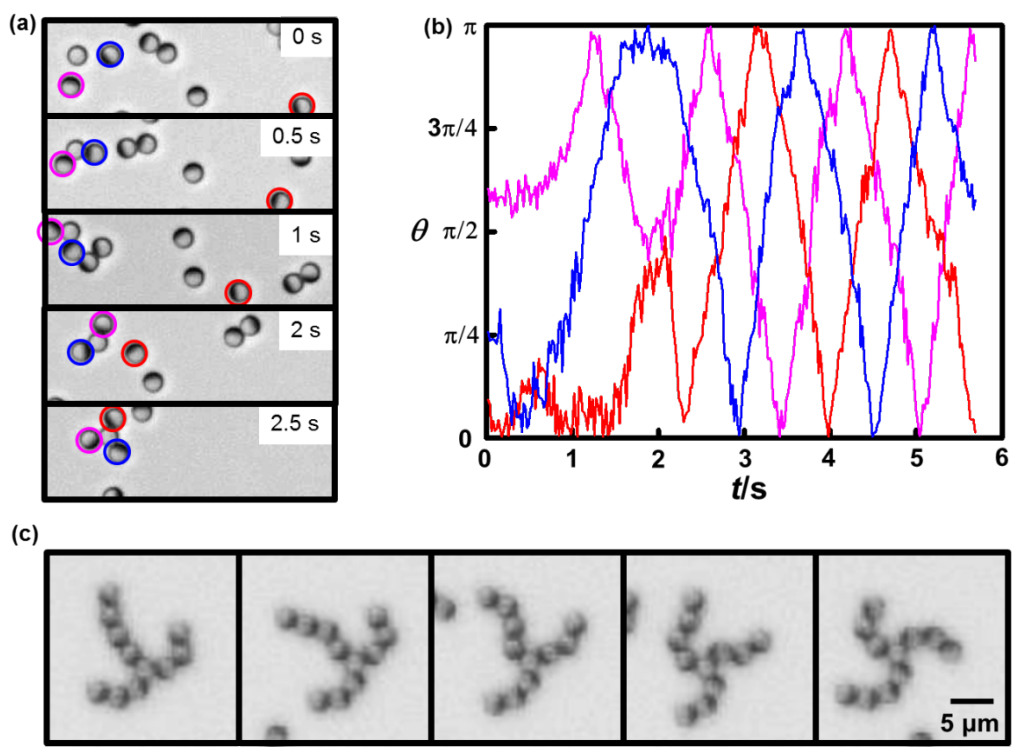

Figure 3. (a) A time sequence of microscope images showing the simultaneous assembly and coordination of a three-arm pinwheel. Blue, red and magenta circles indicate the identity of participating particles. (b) History of orientation for the three labeled particles in (a). Starting at about 2 s elapsed time, the three active particles coordinate their motion and have fixed orientation differences. Noise of these curves reflects jiggling of particle orientation. (c) The coordination process of a pinwheel with longer arms over two seconds.

Empirically, we find that the relative orientation of neighboring Janus particles in each arm favors an angle around $\pi/4$, not the value of 0 that might be anticipated solely from considering induced dipole interactions. The observed distribution of orientation angle is plotted in Fig. 4. The discrepancy is believed likely to stem from hydrodynamic interactions, but no quantitative explanation is offered at this time. Hydrodynamics probably also promotes the curved shape adopted by rotating filaments, but if we assume the interaction to be local, defects such as the kink in the middle of the free filament in the inset of Fig. 4 cannot be avoided. The arms in rotating pinwheels almost always lack kinks, however.

Kink-free filaments, which are required for the sustainable rotation of a pinwheel, present a result of "evolution" and "natural selection" in this colloidal world. Here "evolution" means the dynamic assembly and disassembly of all possible structures through the various "reactions" that happen in the system. Long-lived structures are "naturally selected" because the combined collective motion of the clusters is compatible with each active individual particle inside.

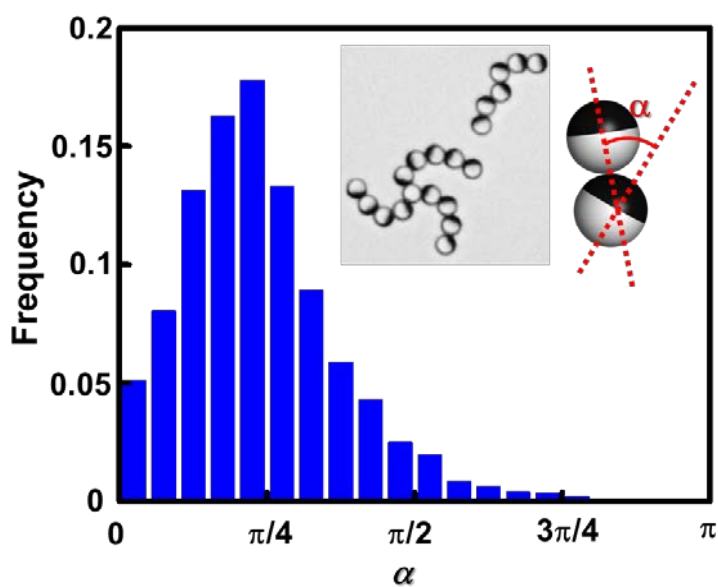

Figure 4. The distribution of the orientation difference between two adjoining Janus particles, from analyzing chains on rotating pinwheels and also those that form free chains. Inset illustrates the coexistence of free and confined active filaments. The dataset includes about 15,000 pairs of particles over 20 seconds

### *Cluster reactions*

Imaging the malleability and reconfigurability of this system, we find that this directed dynamics falls into categories that are familiar in organic chemistry: addition, substitution, ring opening and closure, as well as more complex reactions. Fig. 5 shows four typical reactions. In Fig. 5a, the magenta particle approaches the blue particle at the end of an arm on a rotating pinwheel, quickly adjusts its orientation and attaches. Given the requisite distance and mutual orientation between particles, this reaction can also occur internally within a cluster, as illustrated in Fig. 5b, where by chance the blue particle at the end of one arm lags to the point that it "reacts" with the nearby magenta particle on the neighboring arm: the orientation is acceptable and distance sufficiently close, so bonding between blue and magenta particles is created. Ring closure does not necessarily change the rotation of the overall cluster, because each particle is propelled by its own driving force according to its location within the cluster. In Fig. 5c, the cyan particle approaches the blue and magenta particles and inserts between them. Alternatively, this can be interpreted to show that the cyan particle substitutes the magenta particle in bonding with the blue particle. During the transition state the three particles form a triangle and bondings between them shuffle. Fig. 5d illustrates more complex reactions: with the blue particle attacking the magenta particle, the magenta-yellow bonding breaks, which leads to further detachment of the cyan-yellow dimer.

Superficially-similar colloidal "reactions" might be observed in equilibrium self-assembly.[22] The difference here is that the reactions are biased towards the directed evolution of sustainable shapes.

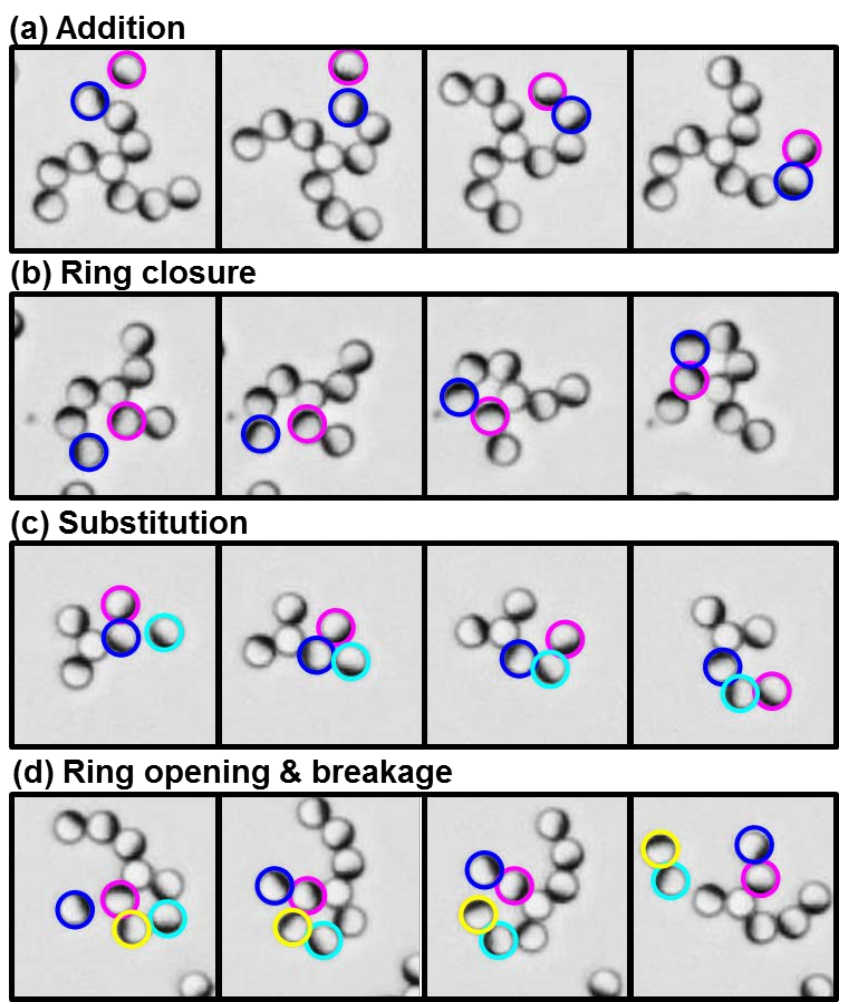

Figure 5. Four examples of typical cluster "reactions". (a) The free magenta particle attaches to the blue particle on a pinwheel during its steady rotation. (b) The blue particle on one arm attaches to the magenta particle on a neighboring arm of the same pinwheel but the process imposes minimal influence on the rotation of the pinwheel. (c) The cyan particle approaches the bonded blue and magenta particles and inserts between them. (d) The blue particle approaches the magenta particle to open the ring bonding between magenta and yellow particles. The yellow and cyan dimer leaves the cluster due to this unfavorable orientation change.

### _Chirality switch_

We can define a state number to identify the rotation condition and structure of particles and clusters: 0 for the state of random motion without chirality, 1 and -1 for clockwise and counterclockwise rotation, respectively. The natural selection process analyzed in this study concerns how to adapt the initial particle assembly (state 0) into rotating pinwheel with either 1 or -1 state. To switch chirality of a cluster requires the collective change of almost all particle orientations.

To switch the chirality of a cluster which rotates already, one mechanism is through complicated reactions that may involve addition, substitution, ring closure and opening. The connectivity of particles within the cluster need not change, but usually does: this usually involves at least one particle that is not part of the cluster originally. Fig. 6a provides one example in which the reactions reconfigure the cluster and switch the chirality from -1 to 1 without changing the connectivity of the three particles immediately adjacent to the centre particle. The cluster starts with three arms of one-, one- and two-particle length respectively, rotating counterclockwise, as shown in the first two frames. The chirality switch occurs between frames 2 and 3, when an outside cyan particle approaches randomly nearby, attaches to the red particle, and induces dramatic changes of the orientation of red and magenta particles to favor rotation in the opposite direction. The reaction, however, does not end there. The close distance and compatible orientation of the yellow particle on one arm and the red particle on the neighboring arm drives them to stick and to continue with clockwise rotation. The small size of this ring and the critical orientation difference between neighboring particles results, with further rotation of this cluster, in detachment from the cluster of the cyan-magenta dimer.

The concept of VIP, very important particle, is illustrated by the development history of this cluster. We can see that the three particles (blue, red, and magenta) that originally form the skeleton of the rotating cluster have always been the VIPs of this particle collection and that the deviation angle of the head of these particles defines the cluster motion. Other particles, whether present as an extension of the blue VIP (yellow particle in frame 1) or as a connection between two VIPs (such as yellow and cyan in frame 5), may bond or debond without affecting significantly the cluster rotation, provided the VIPs do not change their orientations much.

Arms longer than 3 particle diameters can be very flexible as the persistence length of the assembled chains is quite short, as a result of the preferred angle difference between neighboring Janus particles, so the persistence length of assembled chains is quite short. As a result, another way to switch chirality transpires internally but only when the spiral's arm is long enough to be flexible. Although the VIPs immediately adjacent to the centre determine the motion of the overall cluster, other particles on a long arm may affect the orientation of VIPs and eventually change the structure and rotation of the cluster. In Fig. 6b, an example of a 4-arm pinwheel with 3, 3, 3, and 5 arm lengths is presented: the longest arm was unable to keep the original form, deformed, and dragged the other three arms to rotate to the opposite direction. This may have been induced purely internally, enabled by the freedom of bonding between particles on the same arm, but it may have been influenced by interaction with a nearby cluster. The same can happen with a three-arm pinwheel when one arm is sufficiently long.

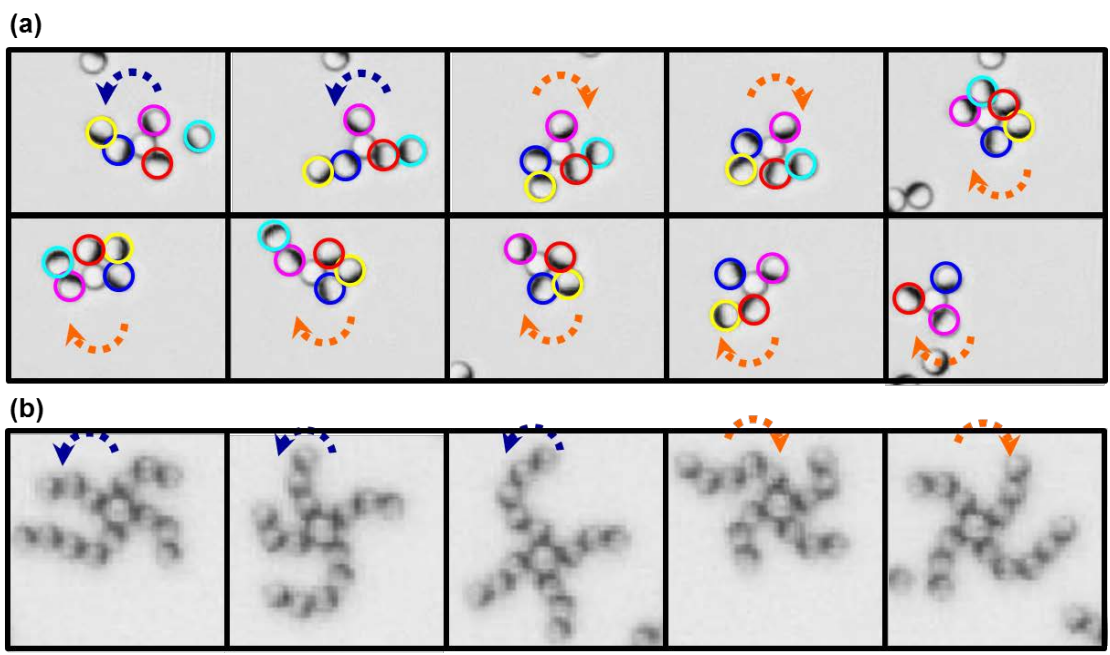

Figure 6. Two ways to switch chirality. (a) After complicated reactions involving emergence and breakage of bonds, the connectivity of the three particles immediately adjacent to the centre particle stayed the same but the chirality changed, in this illustration. (b) With a long enough arm, lack of control of the arm end by the centre particle self-induced the chirality change, in this example.

### *Cluster-cluster interactions*

It has been shown that for arms on a rotating pinwheel the rotation coordination is likely established via mediation by the centre particle. But clusters that are able to translate in space have opportunities for additional interaction. In addition to such communication "cabled" by physical contact, it might also be "wireless" through the longer-range medium of hydrodynamic interactions and induced dipole interactions, each of which could also perturb particle orientations, because as shown in the previous section, small perturbations (internal and external) to the mutual particle orientation within a cluster may result in cascading conformation changes. Physically, one may anticipate intuitively that it would be favorable for nearby clusters to synchronize in the rotation direction that minimizes their mutual influence. Fig. 7a shows one example of synchronization between two nearby clusters, with the left one developing from a translating state. Arrows show direction of translation or rotation for the two clusters. Fig. 7b plots the distance between the two cluster centres with time: it first decreases dramatically due to the translation of the left cluster and then goes up and down as a result of the reconfiguration. Nevertheless, it remains to be determined from a more extensive statistical study how much of conformation change can be attributed strictly to external influence by other clusters, how much of it to purely internal stochastic perturbations.

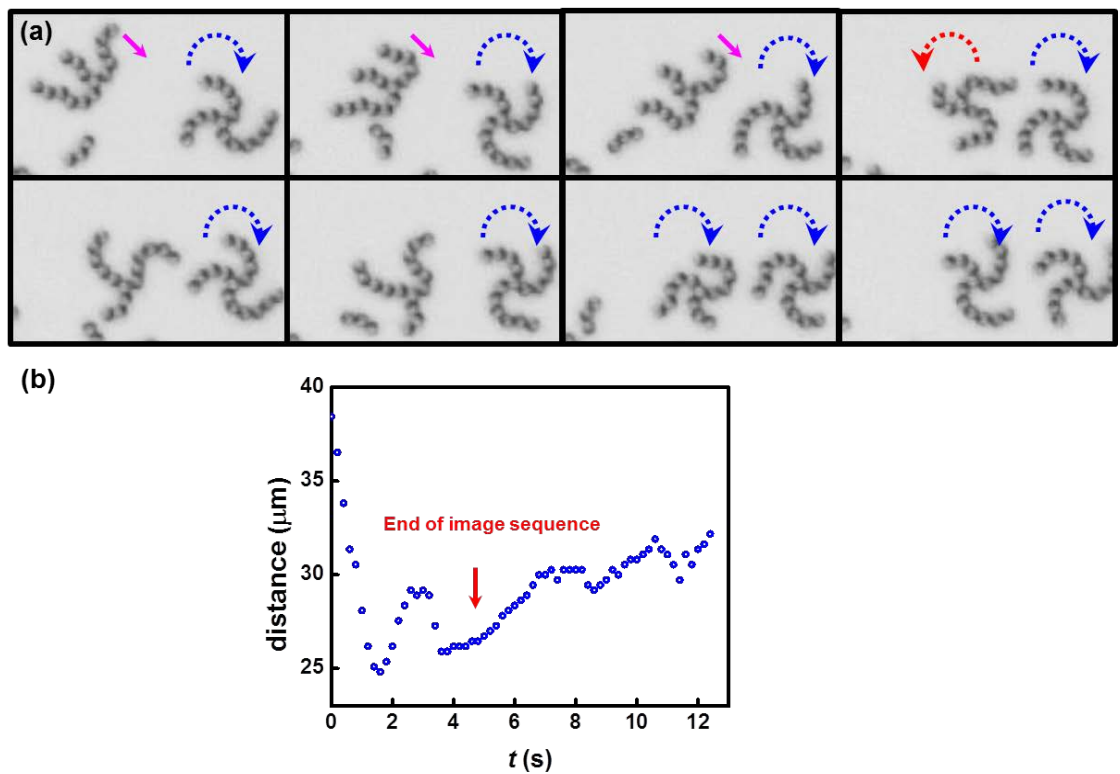

Figure 7. (a) Time sequence of microscopic images showing the synchronization of two nearby clusters over five seconds. The cluster on the right maintains steady rotation through this time but the cluster on the left starts with directional motion towards the one on the right, goes through the indicated adaptation process, and finishes with rotation in the same direction as the other one. Arrows indicate direction of motion or rotation. (b) Plot of distance between the two cluster centres with time. Red arrow points to where the image sequence ends.

## *Conclusion*

Natural selection, normally associated with the differential survival advantage of competing organisms, also applies to the world of active matter, where only certain self-assembled motile structures are compatible with long lifetime. This produces biased "evolution" towards those structures. We have studied this problem in a system where each moving element is visualized in real time, a water suspension of half-metal Janus spheres mixed with a minority of silica beads and subjected to AC voltage that produces not only active motion from induced charge electrophoresis but also particle-particle attraction from induced dipole moments. We observe chiral spirals which rotate clockwise or counterclockwise. Mechanisms of spiral formation and of chirality switch are analyzed. Small perturbations to the mutual particle orientation within a given cluster, whether from random internal fluctuations or from interactions with nearby clusters, may result in cascading conformation changes.

## Acknowledgements

This work was supported at the University of Illinois by the US Department of Energy, Division of Materials Science, under award DE-FG02-07ER46471 through the Frederick Seitz Materials Research Laboratory. At the IBS Centre for Soft and Living Matter, SG acknowledges support by the Institute for Basic Science, project code IBS-R020-D1.

1.      M. C. Marchetti, J. F. Joanny, S. Ramaswamy, T. B. Liverpool, J. Prost, M. Rao and R. A. Simha, *Rev. Mod. Phys.*, 2013, **85**, 1143.

2.      W. Wang, W. T. Duan, S. Ahmed, A. Sen and T. E. Mallouk, *Acc. Chem. Res*, 2015, **48**, 1938-1946.

3.      C. Bechinger, R. Di Leonardo, H. Lowen, C. Reichhardt, G. Volpe and G. Volpe, *arXiv:1602.00081*, 2016.

4.      W. C. K. Poon, *arXiv:1306.4799*, 2013.

5.      W. Wang, W. T. Duan, S. Ahmed, T. E. Mallouk and A. Sen, *Nano Today*, 2013, **8**, 531-554.

6.      W. Vanmegen and S. M. Underwood, *Phys. Rev. E*, 1993, **47**, 248-261.

7.      J. C. Crocker, M. T. Valentine, E. R. Weeks, T. Gisler, P. D. Kaplan, A. G. Yodh and D. A. Weitz, *Phys. Rev. Lett.*, 2000, **85**, 888-891.

8.      J. Yan, K. Chaudhary, S. C. Bae, J. A. Lewis and S. Granick, *Nat. Commun.*, 2013, **4**, 1516.

9.      Q. Chen, S. C. Bae and S. Granick, *Nature*, 2011, **469**, 381-384.

10.     Q. Chen, J. K. Whitmer, S. Jiang, S. C. Bae, E. Luijten and S. Granick, *Science*, 2011, **331**, 199-202.

11.     J. Zhang, E. Luijten and S. Granick, *Annu. Rev. Phys. Chem.*, 2015, **66**, 581-600.

12.     H. S. Jennings, *Am. Nat.*, 1901, **35**, 369-378.

13.     L. Hong, A. Cacciuto, E. Luijten and S. Granick, *Nano Lett.*, 2006, **6**, 2510-2514.

14.     S. Jiang, J. Yan, J. K. Whitmer, S. M. Anthony, E. Luijten and S. Granick, *Phys. Rev. Lett.*, 2014, **112**, 218301.

15.     C. Q. Yu, J. Zhang and S. Granick, *Angew. Chem. Int. Ed.*, 2014, **53**, 4364-4367.

16.     A. W. Long, J. Zhang, S. Granick and A. L. Ferguson, *Soft Matter*, 2015, **11**, 8141-8153.

17.     J. Zhang, J. Yan and S. Granick, *Angew. Chem. Int. Ed.*, 2016, **55**, 1-5.

18.     F. D. Ma, S. J. Wang, D. T. Wu and N. Wu, *Proc. Natl. Acad. Sci.*, 2015, **112**, 6307-6312.

19.     F. D. Ma, D. T. Wu and N. Wu, *J. Am. Chem. Soc.*, 2013, **135**, 7839-7842.

20.     R. Suzuki, H. R. Jiang and M. Sano, *arXiv:1104.5607*, 2011.

21.     S. Capozziello and A. Lattanzi, *Chirality*, 2006, **18**, 17-23.

22.     Y. F. Wang, Y. Wang, D. R. Breed, V. N. Manoharan, L. Feng, A. D. Hollingsworth, M. Weck and D. J. Pine, *Nature*, 2012, **491**, 51-55.